\documentclass[prd,twocolumn]{revtex4}
\usepackage{amsmath}
\usepackage{latexsym}
\usepackage{amsfonts}
\newcommand{\be}{\begin{equation}}
\newcommand{\ee}{\end{equation}}
\begin{document}
\title{Extended Holographic dark energy}
\author{Yungui Gong} \email{gongyg@cqupt.edu.cn}
\affiliation{Institute of Applied Physics and College of
Electronic Engineering, Chongqing University of Posts and
Telecommunications, Chongqing 400065, China}
\begin{abstract}
The idea of relating the infrared and ultraviolet cutoffs is
applied to Brans-Dicke theory of gravitation. We find that the
Hubble scale or the particle horizon as the infrared cutoff will
not give accelerating expansion. The dynamical cosmological
constant with the event horizon as the infrared cutoff is a viable
dark energy model.
\end{abstract}
\maketitle

\parindent=4ex

The Type Ia supernova (SN Ia) observations suggest that the
expansion of our universe is accelerating and dark energy
contributes 2/3 to the critical density of the present universe
\cite{sp99,agr98}. SN Ia observations also provide the evidence of
a decelerated universe in the recent past with the transition
redshift $z_{q=0}\sim 0.5$ \cite{agr,mstagr}. The cosmic
background microwave (CMB) observations support a spatially flat
universe as predicted by the inflationary models
\cite{pdb00,sh00}. The simplest candidate of dark energy is the
cosmological constant. However, the unusual small value of the
cosmological constant leads to the search for dynamical dark
energy models \cite{dark,review}. For a review of dark energy
models, see, for example and references therein \cite{review}.
Cohen, Kaplan and Nelson proposed that for any state in the
Hilbert space with energy $E$, the corresponding Schwarzschild
radius $R_s\sim E$ is less than the infrared (IR) cutoff $L$
\cite{hldark1}. Therefore, the maximum entropy is $S^{3/4}_{\rm
BH}$. Under this assumption, a relationship between the
ultraviolet (UV) cutoff and the infrared cutoff is derived, i.e.,
$8\pi GL^3\rho_\Lambda/3\le L$ \cite{hldark1,lvdark}. So the
holographic cosmological constant is
\begin{equation}
\label{hldark} \rho_\Lambda= 3(8\pi G L^2)^{-1}.
\end{equation}
Hsu found that the holographic cosmological constant model based
on the Hubble scale as IR cutoff won't give an accelerating
universe \cite{hldark2}. Li showed that the holographic dark
energy model based on event horizon gave an accelerating universe,
this model was also found to be consistent with current
observations \cite{hldark3,hldark4}.

Einstein's theory of gravity may not describe gravity at very high
energy. The simplest alternative to general relativity is
Brans-Dicke scalar-tensor theory. The recent interest in
scalar-tensor theories of gravity arises from inflationary
cosmology, supergravity and superstring theory. The dilaton field
appears naturally in the low energy effective bosonic string
theory. Scalar degree of freedom arises also upon compactification
of higher dimensions. In this note, we apply the holographic dark
energy idea to Brans-Dicke cosmology.

The Brans-Dicke Lagrangian in the Jordan frame is given by
\begin{equation}
\label{bdlagr} {\cal L}_{BD}={\sqrt{-g}\over 16\pi}\left[\phi R-
\omega\,g^{\mu\nu} {\partial_\mu\phi\partial_\nu\phi\over
\phi}\right]-{\cal L}_m(\psi,\, g_{\mu\nu}).
\end{equation}
In the Jordan frame, the matter minimally couples to the metric
and there is no interaction between the scalar field $\phi$ and
the matter field $\psi$. Here we work on the Jordan frame so that
test particles follow geodesic motion. The gravitational part of
the above Lagrangian (\ref{bdlagr}) is conformal invariant under
the conformal transformations
\begin{gather*}
\gamma_{\mu\nu}=\Omega^2g_{\mu\nu},\quad \Omega=\phi^\lambda ~~
(\lambda\neq {1\over 2}),\\
\sigma=\phi^{1-2\lambda}, \quad {\bar
\omega}={\omega-6\lambda(\lambda-1)\over (2\lambda-1)^2}.
\end{gather*}
Note that the matter Lagrangian ${\cal L}_m(\psi,\, g_{\mu\nu})$
in Eq. (\ref{bdlagr}) is not conformal invariant under the above
conformal transformations. For the case $\lambda=1/2$, we make the
following transformations:
\begin{equation}
\label{conformala} \gamma_{\mu\nu}=e^{\alpha\sigma}g_{\mu\nu},
\end{equation}
\begin{equation}
\label{conformalb} \phi={8\pi\over \kappa^2}e^{\alpha\sigma},
\end{equation}
where $\kappa^2=8\pi G$, $\alpha=\beta\kappa$, and
$\beta^2=2/(2\omega+3)$. Remember that the Jordan-Brans-Dicke
Lagrangian is not invariant under the above transformations
(\ref{conformala}) and (\ref{conformalb}). The homogeneous and
isotropic Friedmann-Robertson-Walker (FRW) space-time metric is
\begin{equation}
\label{rwcosm} ds^2=-dt^2+a^2(t)\left[{dr^2\over
1-k\,r^2}+r^2\,d\Omega\right].
\end{equation}
Based on the flat FRW metric and the perfect fluid
$T_m^{\mu\nu}=(\rho+ p)\,U^\mu\,U^\nu + p\,g^{\mu\nu}$ as the
matter source, we can get the evolution equations of the universe
from the action (\ref{bdlagr}):
\begin{gather}
\label{jbd1} H^2+H{\dot{\phi}\over \phi}-{\omega\over
6}\left({\dot{\phi} \over \phi}\right)^2={8\pi\over 3\phi}\rho,\\
\label{jbd2} \ddot{\phi}+3H\dot{\phi}=4\pi\beta^2(\rho-3p),
\\
\label{jbd3} \dot{\rho}+3H(\rho+p)=0.
\end{gather}
For ordinary pressureless dust matter, $p_m=0$, we have
$\rho_m\,a^3=\rho_{m0}\,a_0^3$, here subscript $0$ means the
current value. During matter dominated epoch, we can get power-law
solutions to Eqs. (\ref{jbd1}) and (\ref{jbd2}):
\begin{equation}
\label{jbd5} a(t)=a_0\,t^p,\quad \phi(t)=\phi_0\,t^q,
\end{equation}
where
\begin{equation}
\label{jbd6} p={2+2\omega\over 4+3\omega}, \quad q={2\over
4+3\omega},
\end{equation}
and $[q(q-1)+3pq]\phi_0=4\pi\beta^2\rho_{m0}$. We set $t_0=1$.

In Brans-Dicke theory, the scalar field $\phi$ takes the role of
$1/G$, so we propose to modify the holographic dark energy
equation (\ref{hldark}) as \be \label{bddark} \rho_\Lambda=
{3\phi\over 8\pi L^2}. \ee Now let us consider the dark energy
dominated universe. First, we choose $L=H^{-1}$. Substitute the
relation to equations (\ref{bddark}), we can get the solution to
equation (\ref{jbd1}), \be \label{hphi} {\phi\over
\phi_0}=\left({a\over a_0}\right)^{6/\omega}.\ee Combining
equations (\ref{bddark}) and (\ref{hphi}) with equations
(\ref{jbd2}) and (\ref{jbd3}), we can get the following power-law
solutions \begin{gather}
a(t)=a_0 t^{\omega/(4\omega+6)},\\
\phi(t)=\phi_0 t^{3/(2\omega+3)},\\
\rho_\Lambda={3\omega^2\over 8\pi(4\omega+6)^2}\phi_0\left({a\over
a_0}\right)^{-2(4\omega+3)/\omega}. \end{gather} To get
accelerating expansion, we must require $-2<\omega <0$. Even
though the low energy effective theory of the string theory can
lead to $\omega=-1$, the current classical experimental
constraints on $\omega$ is $\omega>500$, so the choice of Hubble
scale as the IR cutoff cannot give an accelerating universe. Next,
we choose the particle horizon as the IR cutoff. The particle
horizon was proposed by Fischler and Susskind to apply the
holographic principle to cosmology \cite{bdhor}. In \cite{gong},
it was also shown that the holographic principle by using the
particle horizon was applicable in Brans-Dicke cosmology. With the
choice of particle horizon, we get
\begin{gather}
L=R_{\rm H}=a(t) \int_0^t {d{\tilde
t}\over a({\tilde t})},\\
\rho_\Lambda= {3\phi\over 8\pi R_{\rm H}^2}. \end{gather}
Substitute this holographic dark energy into equations
(\ref{jbd1})-(\ref{jbd3}) and look for power-law solutions
$a(t)=a_0 t^p$ and $\phi(t)=\phi_0 t^q$. From these power-law
solutions, we get the particle horizon $R_{\rm H}=t/(1-p)$.
Substitute these solutions to equations (\ref{jbd1})-(\ref{jbd3}),
we get
\begin{gather}
\label{rel1} p(q+2)={\omega\over 6}q^2+1,\\
\label{rel2} (2\omega+3)pq(3p+q-1)=(p-1)^2(12p+q-2).
\end{gather}
The solutions to equations (\ref{rel1}) and (\ref{rel2}) are
$p\sim 1/2$ and $q\sim 4/(2\omega+3)$. Therefore the expansion of
the universe is not accelerating. In table \ref{tab1} we list some
numerical solutions of $p$ and $q$.
\begin{table}[htp]
\begin{tabular}{|l|c|c|c|c|c|c|c|} \hline
$\omega$&10&50&100&500&600&800&1000\\\hline
$p$&0.483&0.496&0.498&0.4997&0.4997&0.4998&0.4998\\\hline
$q$&0.148&0.037&0.019&0.004&0.0033&0.0025&0.002\\\hline
\end{tabular}
 \caption{The values of $p$ and $q$ for different
$\omega$} \label{tab1}
\end{table}
From the power-law solutions, it is easy to see that an
accelerating expansion require $p>1$. However the particle horizon
gives $p<1$. Therefore, the choice of particle horizon as the IR
cutoff does not give an accelerating expansion. Finally, we
consider the event horizon as the IR cutoff.
\begin{gather}
L=R_{\rm h}=a(t) \int_t^\infty {d{\tilde
t}\over a({\tilde t})},\\
\rho_\Lambda= {3\phi\over 8\pi R_{\rm h}^2}.
\end{gather}
To solve equations (\ref{jbd1})-(\ref{jbd3}) with
$\rho=\rho_\Lambda$, we assume that $\phi/\phi_0=(a/a_0)^\alpha$.
Substitute this relation into equation (\ref{jbd1}), we get
$${H\over H_0}=\left({a\over a_0}\right)^{c-1},$$
where $c=\sqrt{1+\alpha-\omega\alpha^2/6}$. So
$$\rho_\Lambda={3\phi_0H^2_0\over 8\pi}c^2\left({a\over a_0}\right)^{2c+\alpha-2}.$$
Combining this solution with equations (\ref{jbd2}) and
(\ref{jbd3}), we get the following equation for $\alpha$ \be
\label{alpha} (2\omega+3)\alpha(\alpha+c+2)=3c^2(\alpha+2c+2).\ee
So $\alpha\sim 4/(2\omega+3)$ and $c\sim 1$. The numerical
solutions to equation (\ref{alpha}) for different $\omega$ are
shown in table \ref{tab2}.
\begin{table}[htp]
\begin{tabular}{|l|c|c|c|c|c|c|c|} \hline
$\omega$&10&50&100&500&600&800&1000\\\hline
$c$&1.064&1.013&1.007&1.001&1.001&1.0008&1.0007\\\hline
$\alpha$&0.196&0.0398&0.01996&0.004&0.0033&0.0025&0.002\\\hline
\end{tabular}
 \caption{The values of $c$ and $\alpha$ for different
$\omega$} \label{tab2}
\end{table}
Therefore the event horizon gives an accelerating universe. In
fact this result is expected. Because Brans-Dicke cosmology
becomes standard cosmology when $\omega\rightarrow \infty$. We
know that in standard cosmology the Hubble scale and the particle
horizon do not provide the holographic dark energy, but the event
horizon gives the holographic dark energy which drives the
accelerating expansion of our universe. Therefore, the event
horizon as the IR cutoff should provide the extended holographic
dark energy in Brans-Dicke cosmology.
\begin{acknowledgments}
The author thanks R.G. Cai for helpful discussions. The work is
fully supported by CQUPT under Grants No. A2003-54 and No.
A2004-05.
\end{acknowledgments}

\end{document}